\documentclass[twocolumn,epjc3]{svjour3}  
\smartqed  
\RequirePackage{graphicx}
\usepackage{orcidlink}
\usepackage{xcolor}
\usepackage{hyperref}
\hypersetup{
  colorlinks,
  citecolor=blue,
  linkcolor=blue,
  urlcolor=blue}
\usepackage{amsmath}
\journalname{Eur. Phys. J. C}

\begin{document}

\title{Propagation of light in the presence of gravity generated by static and spherically symmetric curved space-times using Maxwell equations
}


\author{Enderson Falcón-Gómez\orcidlink{0000-0002-0008-0624}\thanksref{affil_1} \and Adrián Amor-Martín\orcidlink{0000-0002-6123-4324}\thanksref{affil_1} \and
Valentín De La Rubia\orcidlink{0000-0002-2894-6813}\thanksref{affil_2} \and
Gabriel Santamaría-Botello\orcidlink{0000-0003-4736-0030}\thanksref{affil_3} \and
Vittorio De Falco\orcidlink{0000-0002-4728-1650}\thanksref{affil_4,affil_5} \and 
Luis Enrique García Muñoz\orcidlink{0000-0002-3619-7859}\thanksref{affil_1,e1}
}

\thankstext{e1}{e-mail: \href{mailto:legarcia@ing.uc3m.es}{legarcia@ing.uc3m.es} (corresponding author)}


\institute{Signal Theory and Communications Department, Universidad Carlos III de Madrid, Leganés, CP 28913 Spain\label{affil_1}
           \and
Departamento de Matemática Aplicada a las TIC, ETSI de Telecomunicación, Universidad Politécnica de Madrid, Madrid, Spain\label{affil_2}
           \and
Department of Electrical, Computer, and Energy Engineering, University of Colorado Boulder, Boulder, CO, USA\label{affil_3} \and
Scuola Superiore Meridionale,  Largo San Marcellino 10, 80138 Napoli, Italy\label{affil_4} \and
Istituto Nazionale di Fisica Nucleare, Sezione di Napoli, Via Cintia 80126 Napoli, Italy \label{affil_5}
}

\date{Received: 22 August 2022 / Accepted: 17 October 2022}

\maketitle

\begin{abstract}
In this manuscript, we present an alternative method for calculating null geodesics in General Static Isotropic Metrics in General Relativity and Extended Theories of Gravity. By applying a conformal transformation, we are able to consider an analogue gravity model, where curvature is encoded in the dielectric and magnetic properties of a medium. In other words, we pass from curved to flat space-times, where instead of the Einstein field equations, the Maxwell equations are solved. Within this geometrical background, the photon geodesics are calculated. Then, given different black hole and wormhole metrics, we apply this method obtaining an excellent agreement with respect to the exact solutions in the original gravity framework by committing angular deviations below $3^{\circ}$. Finally, we provide the image of a Schwarzschild black hole surrounded by a thin accretion disk, and the apparent image of a Morris \& Thorne-like wormhole within an angular discrepancy below $4^{\circ}$.
\end{abstract}

\section{Introduction}

When an electromagnetic (EM) field interacts with a strong gravitational field, its characteristics (such as frequency, amplitude, and propagation direction) are affected by the curvature of the space-time. The great interest in studying the behavior of the EM waves’ propagation in different gravitational fields motivated the efforts of some research groups towards the development of EM-gravity analog models conformally settled in flat space-times in order to investigate gravity through an alternative point of view and with very handy tools, which can be reproducible and built up in laboratory. In order to inquiry the curvature properties of the underlying space-time we study the geodesic orbits followed by both photons and massive bodies when they are in free-fall and only gravity is acting on them, neglecting thus other minor perturbing effects.   

The first idea of this conformal EM-gravity analogue pattern can be traced back to Eddington, who in 1920 thought to describe the light deflection effects from the Sun's gravitational field by treating it as it was a refractive medium \cite{eddington1987space}. This concept was better mathematically worked out in 1960 by Plebanski \cite{plebanski1960electromagnetic}, who presented, for the first time, an EM model of the gravitational field based on a set of suitable dielectric permittivity and magnetic permeability tensors.

There are several international research groups aiming at reducing the computational times for calculating the photon trajectories starting from static and spherically symmetric space-times. These developments permit to inquire gravity and compact object physics, such as black holes (BHs) and wormholes (WHs), via numerical imaging, where theoretical results can be benchmarked with those furnished by the Event Horizon Telescope Collaboration \cite{EHC20191,EHC20194,EHC20195,EHC20196,Akiyama2022image}. 

In 1979 Luminet provided the first ever numerical image of the optical appearance of a Schwarzschild BH surrounded by a thin accretion disk \cite{luminet1979image}. Beloborodov \cite{beloborodov2002gravitational}, La Placa et al. \cite{placa2019approximation}, and Poutanen \cite{Poutanen2020} presented different and very accurate approximated expressions of photon geodesics in the Schwarzschild geometry. Also De Falco and collaborators approximated photon geodesics in Schwarzschild BH, but introducing a mathematical method to obtain polynomials \cite{vittorio2016approximate} and then tested different theories of gravity through ray-tracing procedures \cite{defalco2021testing}. Bao \cite{bao1994emission} and Aldi \cite{aldi2017relativistic} employed geodesics in the estimation of emission lines from the accretion disk around a BH. Johannsen et al. \cite{johannsen2010testingI,johannsen2010testingII} introduced a framework to test the no-hair theorem via electromagnetic observables (i.e., spectrum emission lines from accretion disks, measurement of the Inner Most Stable Orbit (ISCO), and BH images), which they estimated by numerically integrating the geodesics. Bakala et al. \cite{bakala2007extreme} computed the optical appearance of a Schwarzschild-de Sitter BH. Fernández-Núñez et al. \cite{fernandez2016anisotropic} applied a conformal transformation to mimic the Schwarzschild BH space-time with metamaterials. In \cite{falcon2022analogous}, the authors presented and validated a method for computing null geodesics in General Static Isotropic Metrics (GSIMs) by applying a Conformally Flat Space-time Transformation (CFST), based on the Snell's refraction law and isotropic coordinates, particularizing therefore the Plebanski's results to isotropic EM materials.

Regarding the last work, in this paper we aim to extend those goals by applying the aforementioned strategy to different BH and WH metrics by also discussing about the numerical approximation errors and calculating the apparent image of an Ellis WH. Our methodology might represent an interesting tool to understand and study gravity via feasible EM apparati. This article is organized as follows: in Sec. \ref{sec:methods} the theoretical background and the proposed ray-tracing algorithm are presented;
in Sec. \ref{sec:result}, we show the outcomes of our approach; finally in Sec. \ref{sec:conclusion}, we discuss the obtained results and draw the conclusions.

\section{Methods}
\label{sec:methods}

\subsection{The General Static Isotropic Metric}
\label{sec:GSIM}

In General Relativity (GR), the features of a space-time geometry are encoded in the metric tensor $g_{\alpha \beta}$. The GSIM describes static and isotropic gravitational fields, namely, they are independent from the coordinate time $x^0 = t$ and depend only on the radial coordinate \cite{weinberg1972gravitation}. Their line element, $d s^2 = g_{\alpha \beta} dx^{\alpha} dx^{\beta}$ (where Einstein' summation convention is used), expressed in spherical coordinates $(t,r,\theta,\phi)$, geometric units ($c= G =1$, where $c$ is the speed of light in vacuum and $G$ is the gravitational constant), and signature $(-,+,+,+)$ reads as
\begin{equation} \label{eq:generalStaticIsotropicMetricSphericalStandardForm}
ds^2 = - g_{\rm tt}(r) dt^2 + g_{\rm rr}(r)dr^2 + r^2( d\theta ^2 + \sin^2 \theta d\phi^2),
\end{equation}
where $g_{\rm tt}(r)$, $g_{\rm rr}(r)$ are two unknown positive functions (depending only on the radius, $r$), whose explicit expressions allow to determine the background space-time geometry.

\subsection{Photon geodesic equations in GSIM}
\label{sec:photn_geodesics}
In GR, the null geodesic trajectories represent the minimal path connecting two points followed by photons on the space-time background. They can be obtained via the Euler-Lagrange equations {$\frac{\mathrm{d}}{\mathrm{d} \lambda}\left( \frac{\partial \mathcal{L}}{\partial \ \dot{x}^ \alpha}\right)-\frac{\partial \mathcal{L}}{\partial \ x^ \alpha}=~0$}, where $x^\alpha$ are the coordinates, $2\mathcal{L} = g_{\alpha \beta} \frac{dx^\alpha}{d\lambda}\frac{dx^\beta}{d\lambda}$ is the Lagrangian function, and $\lambda$ is the affine parameter along the null trajectories \cite{defalco2019coupling}. Due to the spherical symmetry of the GSIM, the photon energy $E$ and angular momentum $L$ (along any direction) are conserved along the photon orbit. The photon moves in a single \emph{invariant plane} \cite{vittorio2016approximate}, characterized by the initial position and spatial velocity vector of the photon. Without loss of generality, we can settle the photon dynamics in the equatorial plane, i.e., $\theta=\pi/2$. Its orbital equation in a GSIM, derived from the Euler-Lagrange equations, has the following expression
\begin{equation} \label{eq:orbitalequation}
\left( \frac{\partial r}{\partial \phi} \right)^2 = -\frac{r^4}{g_{\rm rr}(r) g_{\rm tt}(r)}\left[ \frac{1}{ b^2}+\frac{g_{\rm tt}(r)}{r^2}\right],
\end{equation}
where $b =L/E$ is the photon impact parameter, which can be defined as follows
\begin{equation} \label{eq:impactparameter}
b = \frac{r \sin(\alpha)}{\sqrt{-g_{\rm tt}(r)}},
\end{equation}
with $\alpha$ being the emission angle. The so-called critical impact parameter, $b_{\rm c}$, is obtained by Eq. (\ref{eq:impactparameter}) substituting $r=r_{\rm ps}$ and $\alpha = \pi/2$, where $r_{\rm ps}$ is the photon-sphere radius (obtained from the effective potential $V_{\rm Eff} = -g_{\rm tt}(r)/r^2$ by imposing $\frac{d V_{\rm Eff}}{dr}=0$) \cite{defalco2021testing}. We can classify three possible families of null-geodesic trajectories depending on the values of $b$ with respect to the critical photon impact parameter $b_c$: (1) $b<b_c$, which include all photons coming from infinity and ending their motion inside the compact object's boundary; (2) $b=b_c$, corresponding to the ``critical orbits'', where the photons from infinity conclude their motions on a stable circular orbit around the compact body, namely the photon sphere $r_{\rm ps}$; (3) $b>b_c$, encompassing all photons coming from infinity, being deflected by the compact object's curvature, and sent again to infinity. As a final remark, not all GSIM-like metric may allow photons to move along critical orbits.


\subsection{Conformally flat spacetime transformation}\label{sec:cfst}

A CFST \cite{weinberg1972gravitation,fernandez2016anisotropic,falcon2022analogous} maps the metric \eqref{eq:generalStaticIsotropicMetricSphericalStandardForm} into another being isotropic, whose explicit expression is
\begin{equation}
ds^2 = -H(\rho)dt^2+J(\rho)({d\rho}^2+{\rho}^2 d\theta ^2 + {\rho} ^2 \sin^2 \theta d\phi^2),
\label{eq:generalStaticIsotropicMetricIsotropicForm}
\end{equation}
where $H(\rho)=-g_{\rm tt}(r(\rho))$, $J(\rho) = r(\rho) ^2/\rho ^2$, and the radial coordinate $r$ changes into a new suitable radial coordinate $\rho$, whose transformation is
\begin{equation}
\rho (r) = C_o \cdot \exp \left({\int\limits_{r}^\infty \frac{\sqrt{g_{rr}(r')}}{r'} \mathrm{d}r'}\right),
\label{eq:rhoInFunctionOfr}
\end{equation}
where $C_o$ is a constant that can be computed knowing that $J(\rho) = 1$ for $r\to\infty$, for the asymptotically flat condition. As a final remark, this conformal transformation applies only to GSIM space-times.

\subsection{Analogous Electromagnetic-gravity Model}
The EM approach for treating the GSIM gravitational fields is based on the \emph{analog-gravity model} derived in \cite{plebanski1960electromagnetic}. Since GSIM-like metrics are diagonal in spherical coordinates, the equivalent dielectric permittivity and magnetic permeability tensors are related to the metric tensor via
\begin{equation}
\frac{\left({\overleftrightarrow \epsilon}\right)^{ij}}{\epsilon_0} = \frac{\left({\overleftrightarrow \mu}\right)^{ij}}{\mu_0} = -\frac{\sqrt{-g}}{g_{00}}g^{ij},
\label{eq:subsection:Constitutive_relationships_for_non-covariant_Maxwell_equations_in_curved_spacetime_C}
\end{equation}
where $g=\det(g_{\mu \nu})$. This formalism can be exploited when the EM field amplitudes are low enough such that their effect on the metric tensor is negligible \cite{plebanski1960electromagnetic}. When the metric is expressed in its isotropic form (\ref{eq:generalStaticIsotropicMetricIsotropicForm}) thanks to the CFST transformation  (\ref{eq:rhoInFunctionOfr}), the dielectric permittivity and magnetic permeability tensors assume the following form
\begin{equation}
\frac{\left({\overleftrightarrow \epsilon}\right)^{ij}}{\epsilon_0} = \frac{\left({\overleftrightarrow \mu}\right)^{ij}}{\mu_0} = n (\rho) ^2\delta_{ij},
\label{eq:subsection:Constitutive_relationships_for_non-covariant_Maxwell_equations_in_curved_spacetime_D}
\end{equation}
where $\delta_{ij}$ is the Kronecker delta and 
\begin{equation}
n (\rho) = \sqrt{\frac{J(\rho)}{H(\rho)}},
\label{eq:refractionIndex01}
\end{equation}
$n (\rho)$ being the isotropic refractive index, which allows to rewrite the metric (\ref{eq:generalStaticIsotropicMetricIsotropicForm}) as \cite{kunz1954propagation}
\begin{equation}
{dt}^2 = {n (\rho)}^2 ({d\rho}^2+{\rho}^2 d\theta ^2 + {\rho} ^2 \sin^2 \theta d\phi^2).
\label{eq:generalStaticIsotropicMetricIsotropicForm3}
\end{equation}
Equation (\ref{eq:generalStaticIsotropicMetricIsotropicForm3}) represents the EM-length line element for calculating the optical distance between two neighboring points for a Transversal-ElectroMagnetic (TEM) mode of propagation.


\subsection{Ray-tracing technique in analogous electromagnetic model spacetimes}\label{sec:raytracingalgorithm}
\begin{figure}[h!]
\centering
\includegraphics[scale = 1]{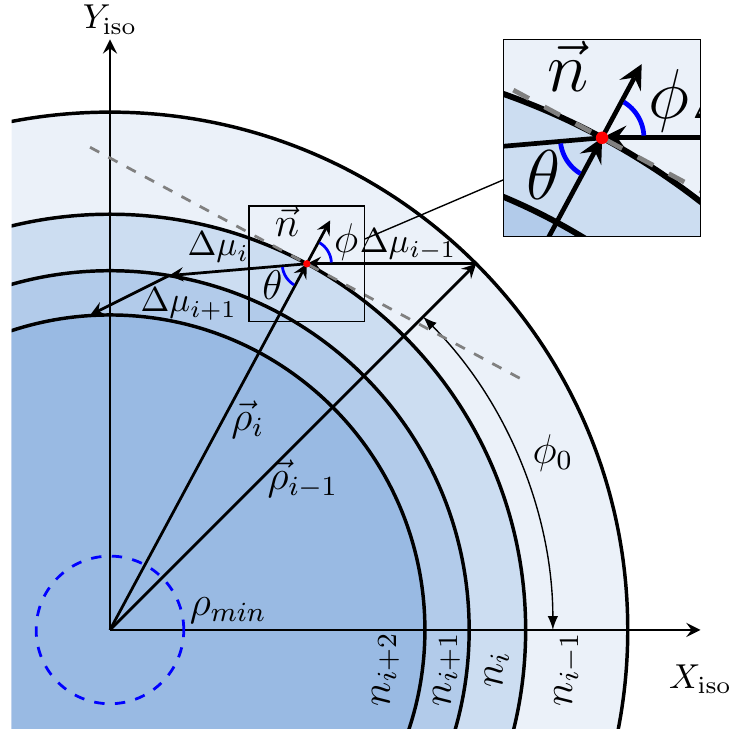}
\caption{Scheme of the method employed to discretize a null geodesic path via straight segments and the Snell's refraction law. The colored circular regions represent a generic GSIM equivalent spacetime described in an isotropic coordinate system through the EM-analog model used in this work. The figure shows how a null geodesic in the inhomogeneous media is approximated by using discrete steps when a photon moves from a region of low refractive index towards a region of high refractive index. }
\label{fig:rayTracingalgorithm}
\end{figure}

We compute the photon geodesics in GSIM-like metrics model
by exploiting the conventional ray-tracing scheme via the Snell's refraction law. In Fig. \ref{fig:rayTracingalgorithm}, we report a visual sketch of the strategy adopted in this article.

\begin{figure}[h!]
\centering
\includegraphics[scale = 1.3]{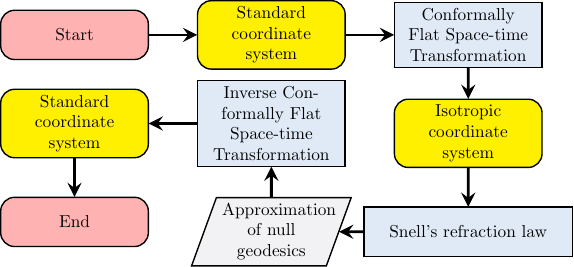}
\caption{Flowchart of the algorithm employed to compute the photon trajectories based on the Snell's refraction law.}
\label{fig:flowchart}
\end{figure}

In Fig. \ref{fig:flowchart}, we sketch the flowchart of the approximation method to calculate photon trajectories in GSIM space-times proposed in this work. Firstly, we set the initial position $(r _{\rm i},\phi _{\rm i})$ and momentum $\vec{\kappa}$ of a photon emitted with an emission angle $\alpha$ in the standard spherical coordinate system or GSIM space-time (see Fig. \ref{fig:flowchart}). The momentum, as a function of the initial conditions, is given by
\begin{equation}
\left( \kappa _{\rm r}, \kappa _{\rm \phi} \right) = \left( \frac{\cos(\alpha)}{\sqrt{g_{\rm rr}(r_{\rm i})}}, \frac{\sin(\alpha)}{r_{\rm i}}\right).
\label{eq:photonmomentum}
\end{equation}
Then, as shown in Fig. \ref{fig:flowchart}, we transform the initial conditions in spherical coordinates: the initial position ($r _{\rm i},\ \phi _{\rm i}$), and the wave vector ($\vec{\kappa} = k_r \vec{e}_r+k_\phi \vec{e}_\phi $ ) to isotropic coordinates: ($\rho _{\rm i},\phi _{\rm i}$), and ($\vec{\kappa} = k_\rho \vec{e}_\rho+k_\phi \vec{e}_\phi $), respectively, by using the Conformally Flat Space-time Transformation. 
The azimuthal angle $\phi _i$ and the azimuthal component of the wave vector are not altered, because the coordinate transformation is conformal. Instead, the radial component of the wave vector transforms as
\begin{equation}
\kappa _{\rm \rho} = \frac{\partial \rho }{\partial r} \kappa _{\rm r} = \frac{\rho \sqrt{g_{\rm rr}(r)}}{r} \kappa _{\rm r}.
\label{eq:transformationVelocity}
\end{equation}
As shown in Fig. \ref{fig:rayTracingalgorithm}, the ray-tracing algorithm can be adapted to any GSIM space-time (\ref{eq:generalStaticIsotropicMetricIsotropicForm3}) considering the movement of photons in an inhomogeneous medium and using the Snell's refraction law, $n_{\rm {i-1}} \sin(\phi) = n_{\rm {i}} \sin(\theta)$, where $\phi$ and $\theta$ are the incidence and refraction angles, respectively, and $n_{\rm {i-1}}$ and $n_{\rm {i}}$ are the refractive indices at the point $i-1$ and $i$, respectively (see inset in Fig. \ref{fig:rayTracingalgorithm}), to compute how the photon's trajectory is refracted because of the medium inhomogeneity. In the following, we shall explain the main characteristics of the solving-technique based on Snell's law. Let us consider a photon moving, such as in Fig. \ref{fig:rayTracingalgorithm}, from a point $i-1$ to a point $i$. We divide the medium into three regions separated by radial distances $\rho_{i-1}$ and $\rho_{i}$, respectively, where, for simplicity, we assume that $\rho_{i}<\rho_{i-1}$ as reported in Fig. \ref{fig:rayTracingalgorithm}. We have that $\rho_{i-1}$ is the initial radial position and $\rho_{i} = \rho_{i}(\Delta \mu _i)$ is the arrival radial position, where $\Delta \mu _i$ is the straight-length followed  by the photon along its wave vector direction. We state that the refractive index inside the region $\rho_{i}<\rho<\rho_{i-1}$ is a constant defined by $n(\rho) = n(\rho _{i-1})$ since we consider very small displacements. 
The photon's movement gives rise to the so-called \emph{electrical displacement} ($\Delta \mu _{\rm ele}$), which provides information on the phase difference between two points, obtained by integrating Eq. (\ref{eq:generalStaticIsotropicMetricIsotropicForm3}) as follows:
\begin{equation}
    \Delta \mu _{\rm ele} = \int\limits_{\lambda_{i-1}}^ {\lambda_i} n (\rho(\lambda)) \sqrt{ \left(\frac{\mathrm{d}\rho}{\mathrm{d}\lambda}\right)^2+{\rho} ^2 \left(\frac{\mathrm{d}\phi}{\mathrm{d}\lambda}\right)^2} \mathrm{d}\lambda,\label{eq:condition_lenght}
\end{equation}
where $\lambda$ is an affine parameter along the photon trajectory, useful for the integration process. Equation (\ref{eq:condition_lenght}) is a line integral along the aforementioned photon straight-line trajectory.
We define $\Delta \mu _{\rm max}$ as the maximum electrical displacement in order to obtain a good resolution and low approximation error (after having fixed an a-priori tolerance). Between two points, the smaller the electrical displacement is, the smaller the variation of the refractive index will be. This ensures that the solving technique based on Snell's law provides a reliable approximation. Therefore, at each step, we check whether the electrical displacement is larger than $\Delta \mu _{\rm max}$, otherwise $\Delta \mu _i$ (defining the photon's length-element along the wave vector direction) is reduced until the condition $\Delta \mu _ {\rm ele} < \Delta \mu _{\rm max}$ is met. After having determined the optimum $\Delta \mu _i$, we can then compute the arrival radial distance $\rho _i$ and the related refractive index in that point, $n(\rho) = n(\rho_{i})$. We calculate the refracted wave vector to be used in the next photon displacement by invoking the Snell's refraction law. This process continues until the photon reaches a maximum number of displacements $N$ or when the photon reaches the spatial infinity, namely the limits of the computational domain. The adaptability of the displacement distance $\Delta \mu _i$ at each propagation step is mandatory provided that the isotropic refractive index, $n(\rho)$, is non-homogeneous and the optimum $\Delta \mu _i$ is not known at the beginning of the displacement where it is set to a default value. Once the approximation finishes, the Conformally Flat Space-time Transformation is inverted to come back to the standard (spherical) coordinate system.

This algorithm must be slightly revised for WHs, since the photons can traverse them. Indeed, care must be encompassed during the integration process on the WH throat (boundary between universe 1 and universe 2). In this region, we use as initial position and four-velocity for the universe 2 the arrival coordinates obtained in the universe 1, together with the radial component of four-velocity inverted, while maintaining fixed the angular component. Physically, this corresponds to the angular momentum conservation. To compute the arrival coordinates, we allow the photon to move towards the origin of the universe 1 closer to $\rho \approx 1.01  \rho_{\rm min}$ (we set this condition heuristically, and it can adjusted in order to improve the precision), which can dub this point the ``pre-arrival'' point. Then, we compute the arrival coordinates by extending the pre-arrival coordinates to the WH throat by intersecting a straight trajectory along the pre-arrival wave vector with the WH mouth. This approximation may give rise to high deviations when going from one universe to another if the refractive index has strong variation near the WH mouth. This way of approximating null geodesics works very well for all kinds of null geodesic in a GSIM (see Sec.~\ref{sec:photn_geodesics}). However, due to the discrete nature of the algorithm and instability of critical geodesics, particular attention must be used in the calculations of these orbits.

\begin{figure*}[h!]
\centering
\includegraphics[scale = 0.65]{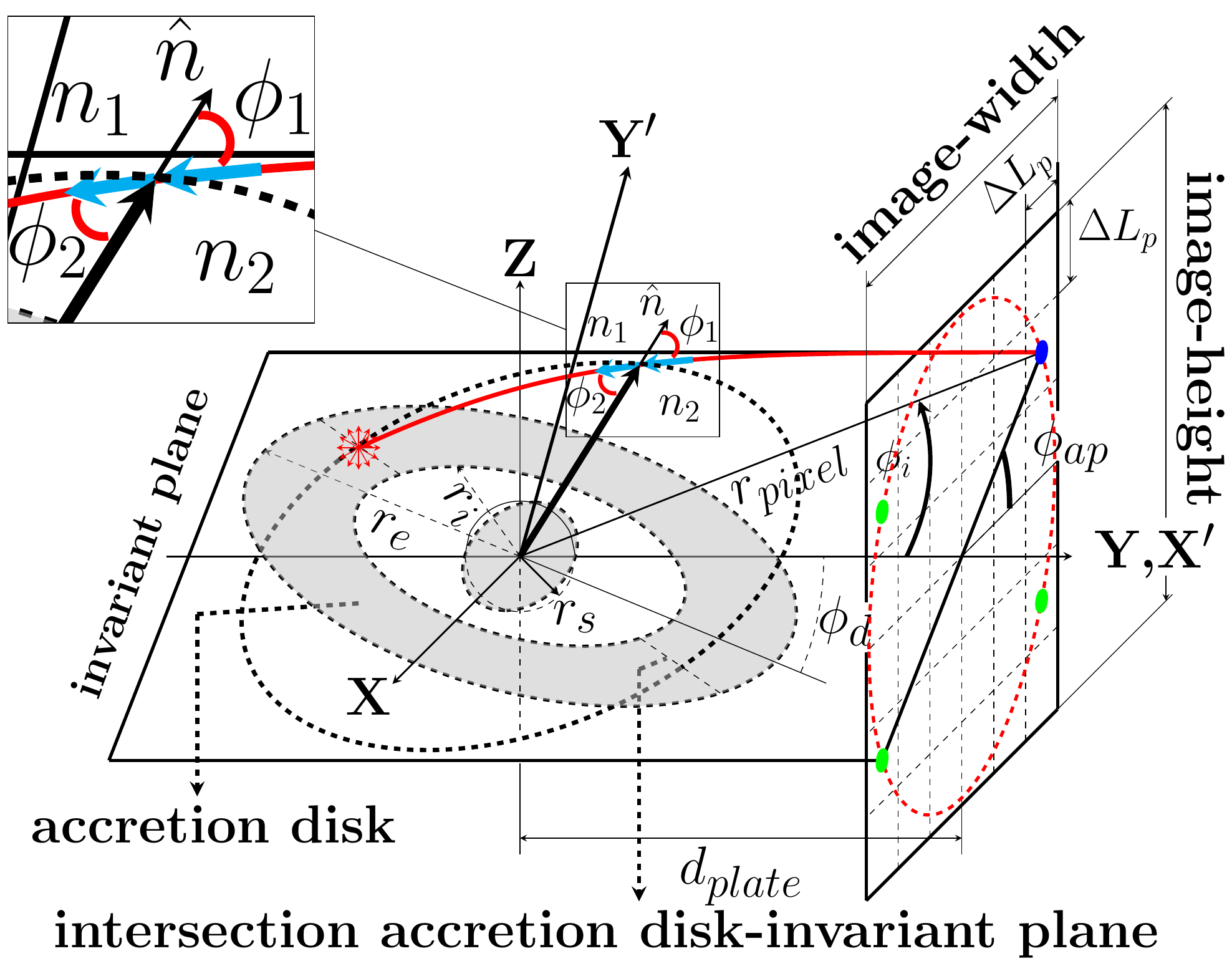}
\caption{Sketch of image generation of a thin accretion disk around a BH. We define $r_{\rm min}\equiv r_{\rm s}=2M$ as the Schwarzschild radius, $r_{\rm i}$ and $r_{\rm e}$ are the inner and outer radii of the accretion disk, respectively. $d_{\rm plate}$ is the distance between the compact object and the center of the photographic plate, $\Delta L_p $ is the pixel squared-size, $r_{\rm pixel}$ is the distance between the pixel center and the origin of the observer's plate, $\Phi _{\rm ap}$ is the rotation angle of the invariant plane that passes through the pixel center around the $Y$ axis, and $\Phi _{\rm d}$ is the rotation angle of the accretion disk around the $X$ axis. The red trajectory is a null geodesic computed in the Schwarzschild metric impinging the accretion disk in a point, which is emitted over there isotropically. $n_{\rm 1}$ and $n_{\rm 2}$ are two refraction indices, $\phi _{\rm 1}$ and $\phi _{\rm 2}$ are the incident and the refraction angles, respectively. $\hat{n}$ is the normal vector to the circles located in the invariant plane. The computational domain is set in the \emph{celestial sphere}. This figure is taken from Ref. \cite{falcon2022analogous}.}
\label{fig:rayTracingFigure}
\end{figure*}

 As already mentioned in Sec. \ref{sec:photn_geodesics}, critical orbits in GSIM-like metric spiral around the WH mouth or BH event horizon stabilizing their motion around the circular photon orbit. In this approach, 
the photon critical orbits are difficult to be approximated, because they can easily deviate from their trajectories when reach the photon sphere, entailing thus large errors. Therefore, in order to avoid such an effect, we implement an analytical extension to the algorithm, which approximates the final part of the orbit (i.e., when the photons start spiralling around the photon sphere) through a circular trajectory.



\subsection{Image generation of compact objects in curved spacetimes}

In this section, we show how to produce the image of a compact object described by a GSIM framed in an analogue EM-gravity model. In Fig. \ref{fig:rayTracingFigure}, we depict the geometrical setup for imaging a thin accretion disk formed around a static and spherically symmetric compact object.

We consider that each element of the accretion disk emits X-ray EM radiation isotropically in its rest frame. Normally, we should compute an infinite number of photon trajectories from all the emission points coming from each geometrical disk element 
and then find those that reach the observer's photographic plate located at infinity. We follow instead the inverse procedure, consisting in shooting the photons from the observer's location towards the emission point . 
We make use of a collision detection algorithm to know where a light ray is being emitted from. The photon trajectories are colored according to the element they come from.

The photographic plate is divided into squared pixels of lengths $\Delta L_p \ll r_{\rm min}$, and it is placed at a distance $d_{\rm plate}$ from the center of the compact object such that the gravitational effects are negligible\footnote{The distance observer--compact object is normally finite, but since it is enormous with respect to the typical size of the gravitational source, it is normally considered infinite in the theory for easing the ensuing calculations.}. In practice, this corresponds to have $|g_{\rm {rr}}-1|<6\%$ (i.e, asymptotic flatness condition set in our simulations). The protocol for launching the photons relies on shooting them towards the compact body from the center of each pixel with normal direction. Since we are in the invariant plane ($X'-Y'$), as reported in Fig.  \ref{fig:rayTracingFigure}, we must consider only the initial radius and azimuthal velocity components on the invariant plane under consideration given by
\begin{equation}
\vec{\kappa} = (\kappa^r,\kappa^{\phi}) = \left(-\cos(\phi_{\rm i}),\frac{\sin(\phi_{\rm i})}{r}\right). 
\label{eq:kvector}
\end{equation}

In order to speed up the integration process we exploit the spherical symmetry of the problem and the fact that photons are launched with a direction perpendicular to the photographic plate. If we look at Fig. \ref{fig:rayTracingFigure}, we can see the point the photon is being launched from to follow the red-color geodesic high-listed as a blue dot.  If we rotate this point on the photographic plate around the Y axis, we will find out that there are at least three more pixels on the photographic plate (high-listed in green), at which photons will be launched with the same initial conditions in the local coordinate system on the the corresponding invariant plane $X', Y'$ (i.e., $\phi _i, r_{\rm pixel}$, and $\alpha _i$). The geodesics at green points can be obtained from the geodesic at the blue point just by applying a rotation of coordinate system.


\section{Results}\label{sec:result}
\begin{figure}[h!]
\centering
\includegraphics[scale = 0.7]{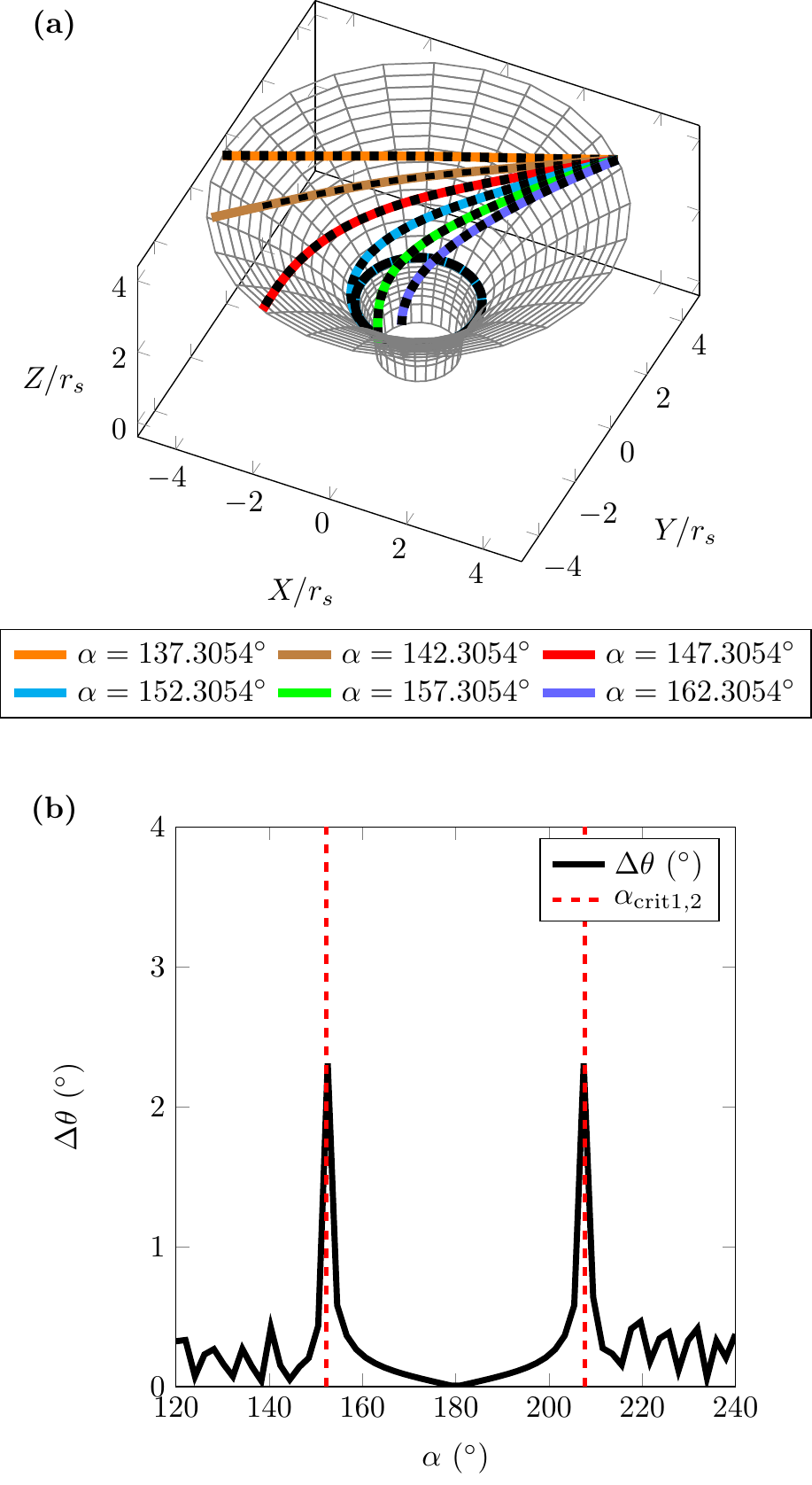}
\caption{Simulation results for Schwarzschild BH case: $r_s = 1$, $R_i = 5r_s$, $\phi _i = 45^{\circ}$, and $\Delta \mu_{\rm  max}=10^{-2}r_s$. (a) Geodesics generated for different emission angles $\alpha$. Color solid lines are results obtained with the proposed algorithm and black dashed lines are obtained via the orbital equation. (b) Angular deviation versus. emission angle $\alpha$ inside the range: $120-240^{\circ}$.}
\label{fig:geodesics_schwarzschild_black_hole}
\end{figure}

In Fig. \ref{fig:geodesics_schwarzschild_black_hole}, we display the geodesics computed in a space-time generated by a Schwarzschild BH \cite{carroll2019spacetime,fernandez2016anisotropic} with $r_s = 2$. Specifically, Fig. {\ref{fig:geodesics_schwarzschild_black_hole}(a)} shows in an embedding diagram \cite{morris1988wormholes} (gray colored mesh) geodesics obtained following the presented approach (colored solid lines) and exact solutions calculated by numerically solving the orbital equation (dashed black lines) for different initial emission angles. In order to measure the convergence of our approach, we plot the \textit{angular deviation}, $\Delta \theta$, measuring the absolute angular error between the final ends of the geodesics computed via our method and the exact solutions. For Fig. {\ref{fig:geodesics_schwarzschild_black_hole}(a)}, it amounts to $\Delta \theta \leq 0.33^{\circ}$.

Instead in Fig.~{\ref{fig:geodesics_schwarzschild_black_hole}(b)}, we plot the angular deviation when photons are launched in a Schwarzschild BH space-time with emission angles in the range $120-240^{\circ}$. It can be seen that around the critical emission angles $\alpha_{\rm crit 1,2}$ the angular deviation is lower than $2.3^\circ$, and when the emission angle is outside the range $\alpha_{\rm crit 1,2}\pm 2^{\circ}$, $\Delta \theta \leq 1^{\circ}$. We expected this result due to the discrete nature of the algorithm and the instability of the critical orbits and trajectories for photons emitted with impact parameter near to the critical impact parameter, $b_c$. This behavior can be improved by increasing the discretization density. As it can be noted from Fig. {\ref{fig:geodesics_schwarzschild_black_hole}(a)}, photons with emission angles larger than the critical one do not fall into the BH, whereas, for critical orbits (cyan solid line), photons are trapped on the photon sphere region. In Fig. {\ref{fig:geodesics_schwarzschild_black_hole}(a)}, for critical orbits, the numerical approximation is analytically extended using a circular trajectory as explained in Sec. \ref{sec:raytracingalgorithm}, and photons with emission angles lower than the critical one fall into the BH event horizon.

\begin{figure}[h!]
\centering
\includegraphics[scale = 0.7]{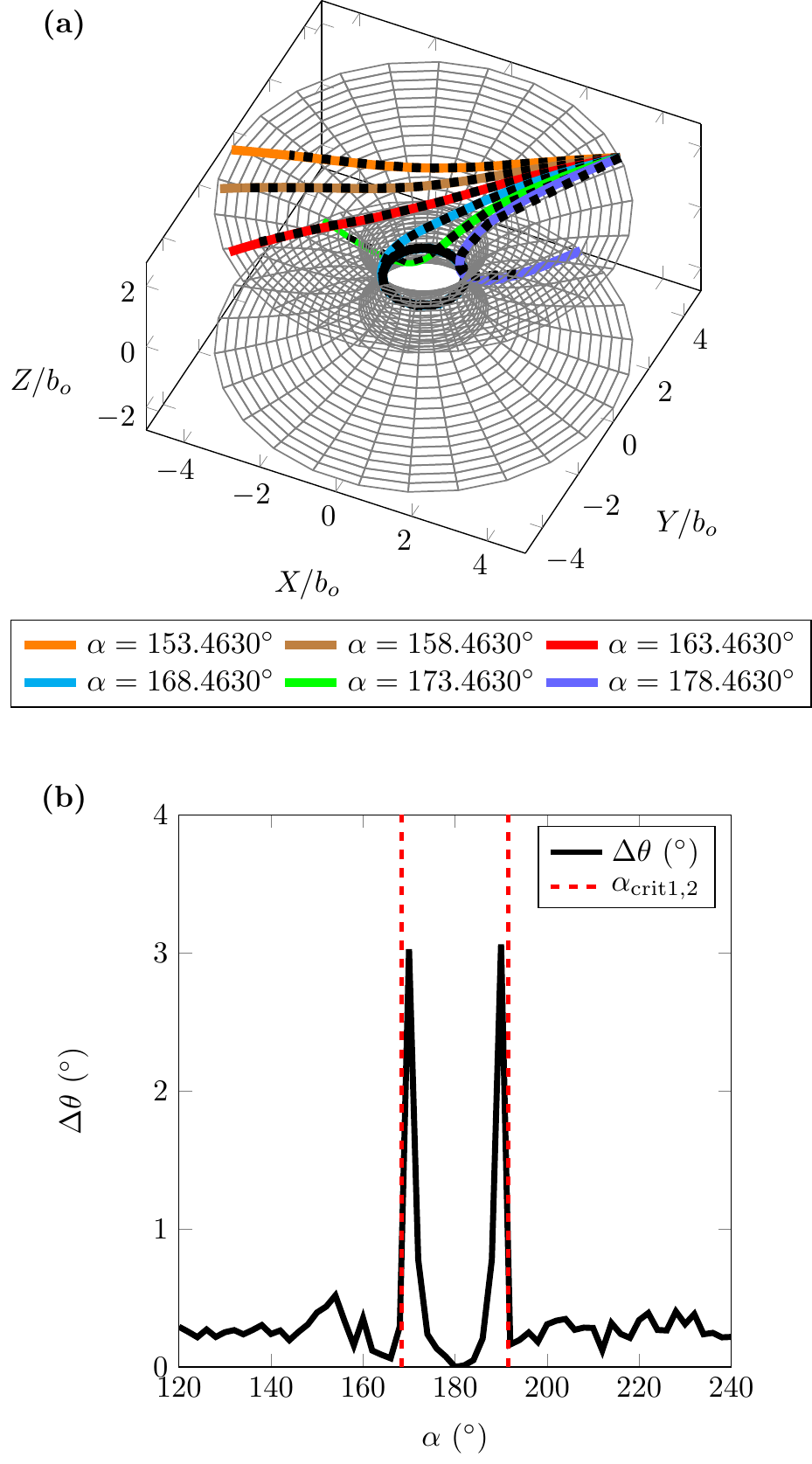}
\caption{Simulation results for an Ellis WH: $b_0 = 1$, $R_i = 5r_s$, $\phi _i = 45^{\circ}$, and $\Delta \mu_{\rm max}=10^{-2}b_0$. Panel (a): Geodesics generated for different emission angles $\alpha$. Solid color are results obtained with the proposed algorithm and black dashed lines are results obtained from the orbital equation. Panel (b): Angular deviation versus emission angle $\alpha$ inside the range: $120^{\circ}-240^{\circ}$.}
\label{fig:geodesics_ellis_wormhole}
\end{figure}

Now we focus our attention on geodesics computed in the Ellis WH space-time \cite{morris1988wormholes} setting the units for the WH throat as $b_0 = 1$, see Fig.~\ref{fig:geodesics_ellis_wormhole}. In particular, Fig. {\ref{fig:geodesics_ellis_wormhole}(a)} shows in the embedding diagram (gray grid) the geodesics computed with our approach (solid lines) and the exact solutions (black dashed lines), having $\Delta \theta \leq 0.63^{\circ}$ in this case. Instead in Fig. {\ref{fig:geodesics_ellis_wormhole}(b)}, we report the angular discrepancy distribution for the emission angle in the range $120-240^{\circ}$. In this case, the maximum $\Delta \theta $ is located around $\alpha_1 = 170^{\circ}$ and $\alpha_2 = 190^{\circ}$ with $\Delta \theta \approx 3^\circ$. Outside the range $\alpha_{1,2}\pm 2^{\circ}$, $\Delta \theta \leq 1^{\circ}$.  As shown in Fig. {\ref{fig:geodesics_ellis_wormhole}(a)}, photons emitted with an emission angle larger than the critical one propagates in the same region (universe 1), whereas photons with emission angle shorter than the critical one go through the WH mouth into the so-called universe 2. The photon sphere of the Ellis WH coincides exactly with the WH mouth. Photons emitted with a critical emission angle (cyan solid line) are trapped in the photon-sphere.

\begin{figure}[h!]
\centering
\includegraphics[scale = 0.7]{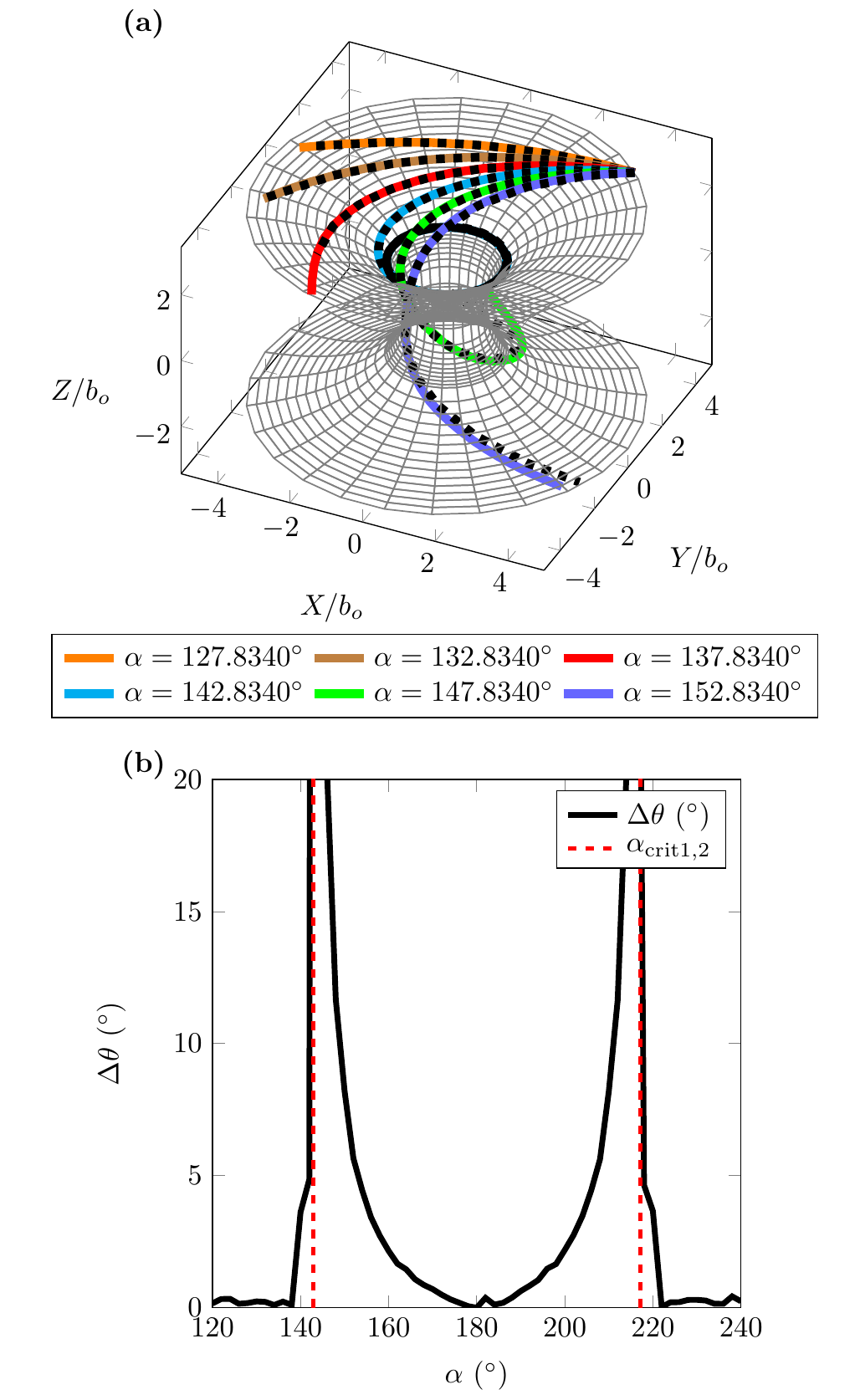}
\caption{simulation results for WH framed in the metric theory of gravity: $(b_0,\gamma) = (0.5,1.5)$, $R_i = 5r_s$, $\phi _i = 45^{\circ}$, and $\Delta \mu_{\rm max}=10^{-2}b_0$. Panel (a): Geodesics generated for different emission angles $\alpha$. Solid color are results obtained with the proposed algorithm and black dashed lines are results obtained from the orbital equation. Panel (b): Angular deviation versus emission angle $\alpha$ inside the range: $120^{\circ}-240^{\circ}$.}
\label{fig:geodesics_metrictheory_wormhole}
\end{figure}

We then consider a Morris \& Thorne-like WH solution framed in a metric theory of gravity (see Ref. \cite{defalco2021testing}  and references therein for more details) endowed with $(b_0,\gamma) = (0.5,1.5)$, see Fig. \ref{fig:geodesics_metrictheory_wormhole}. As done for the previous cases and adopting the same color and style conventions, we first show the geodesics obtained via our method and those obtained integrating exactly the orbital equation using $\Delta \theta \leq 12.8^{\circ}$ (see Fig. {\ref{fig:geodesics_metrictheory_wormhole}(a)}). Then, we highlight the angular discrepancy profile for emission angles in the range $120-240^{\circ}$ (see Fig. {\ref{fig:geodesics_metrictheory_wormhole}(b)}). In the last case, the maximum $\Delta \theta $ is located around $\alpha_1 = 144^{\circ}$ and $\alpha_2 = 216^{\circ}$ with maximum deviation of $\Delta \theta \approx 308.51 ^{\circ}$. Outside the range $\alpha_{1,2}\pm 12^{\circ}$, $\Delta \theta \leq 4^{\circ}$. We note that in this situation, the angular deviation is considerably larger than the previous cases. This anomaly is related to the fact that the WH metric admits a singular point near the WH mouth and there the refractive index blows up. 

\begin{figure}[h]
\centering
\includegraphics[scale = 0.4]{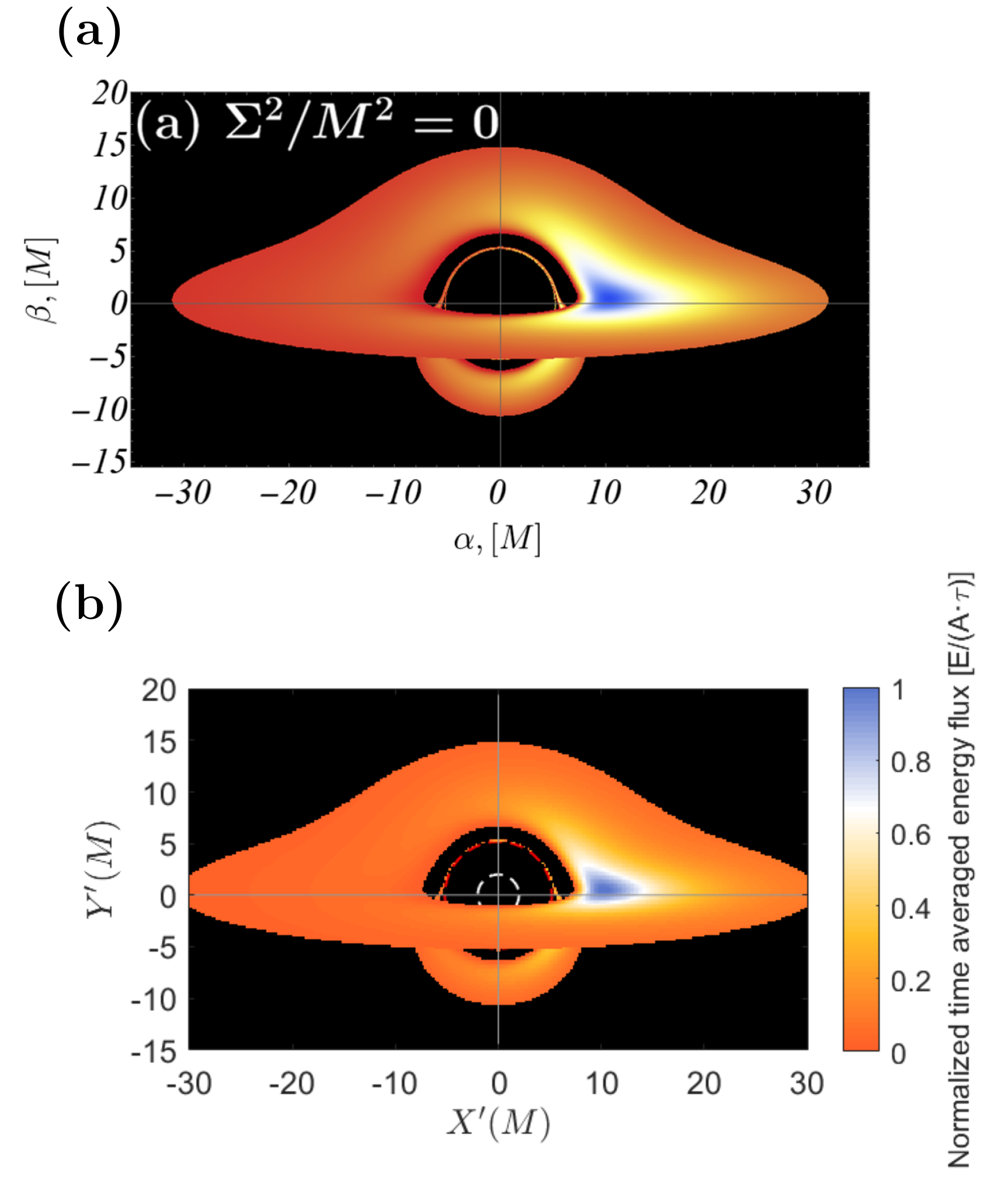}
\vspace{-0.5 cm}
\caption{Apparent image of a Schwarzschild BH surrounded by a thin accretion disk. We set the BH mass $M=1$, $r_s = 2$, $r_{\rm skyph} = 60 r_s$ (computational domain limit or infinite), $\Delta L_{p} = r_s/10$, $d_{\rm plate} = 40 r_s$, $r_i = 3r_s$ (internal radius of the accretion disk), $r_e = 15 r_s$ (external radius of the accretion disk), and $\theta _{\rm o} = 80^\circ$ (inclination angle of the observer). The red and white dashed circles represent the photon sphere radius projected at infinity (BH shadow), and the BH event horizon $r_s$, respectively. The lateral colored scale represents the normalized time averaged energy flux of matter present in the accretion disk. Panel (a): Apparent numerical image of a Schwarzschild BH provided by Igor Bogush \cite{bogush2022photon}, where $\Sigma$ is a coupled scalar field. The vertical and horizontal axes: $\beta$, and $\alpha$, represent the projection of the BH image on the observer's sky. Panel (b): Result obtained with our approach. Credits for textures used in this image to \cite{DNGimages}.}
\label{fig:blackhole_image}
\end{figure}

\begin{figure}[h!]
\centering
\includegraphics[scale = 0.6]{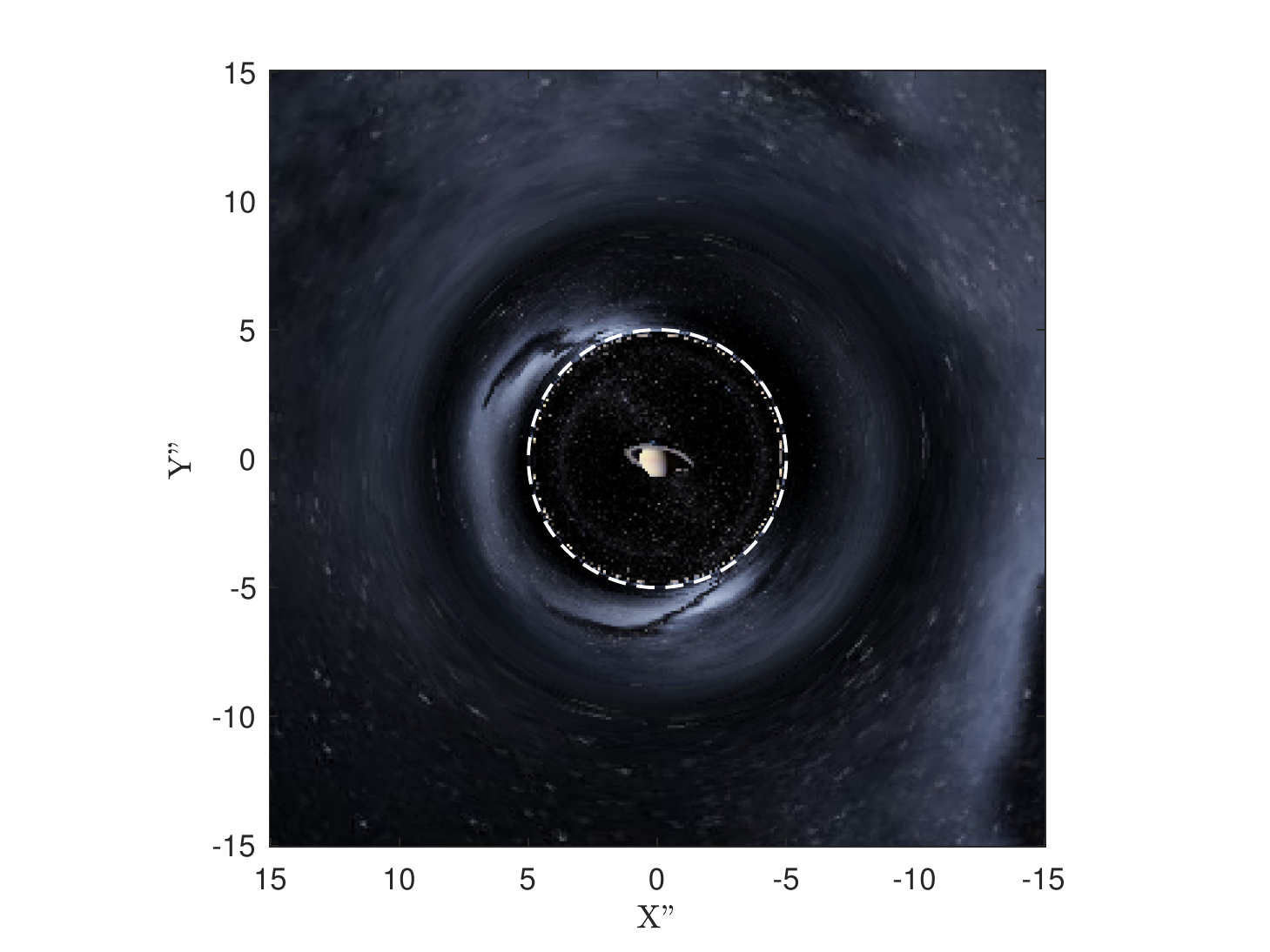}
\caption{Apparent image of a  Morris \& Thorne-like WH in metric theory of gravity \cite{defalco2021testing}. We set $b_0 = 5$, $r_{\rm skyph} = 8 b_0$ (computational domain limit or infinite), $\Delta L_{p} = b_0/50$, and $d_{\rm plate} = 4 b_0$. The white dashed circle represents the WH mouth size $b_0$. Credits for textures used in this image to \cite{DNGimages}.}
\label{fig:wormhole_image}
\end{figure}

Finally, we compare the numerical simulation of an accretion disk around a BH provided by Igor Bogush \cite{bogush2022photon} (see Fig. \ref{fig:blackhole_image}(a)) with the one produced employing our approach (see Fig. \ref{fig:blackhole_image}(b)). The colors of the accretion disk in Fig. \ref{fig:blackhole_image}(b) refer to the observed bolometric flux, as implemented in  \cite{luminet1979image} and \cite{page1974disk}. This numerical simulation shows the characteristic gravitational redshift effects, where matter moving away from the observer is redshifted, whereas that approaching the observer is blueshifted. In addition, the BH shadow (black circular spot surrounded by the BH accretion disk), as seen from infinity, agrees with the theoretical prediction  (red dashed circle with radius $r_{\rm{sh}} = {3 \sqrt{3} r_{\rm{s}}}/{2}$) as reported in \cite{perlick2022calculating}. 
 In Fig. \ref{fig:wormhole_image}, we concentrate on the apparent image of an Ellis WH with $b_0=5$ as seen from the universe 1, where the observer is located. We can see, outside the WH mouth (white dashed circle) how the space-time is distorted because of gravity, whereas inside the mouth, a view of the universe 2 appears.\vspace{-0.4 cm}
\vspace{-0.3 cm}
\section{Conclusion}\label{sec:conclusion}
The null geodesic trajectories, calculated by exploiting the proposed algorithm shown in Figs. \ref{fig:geodesics_schwarzschild_black_hole}, \ref{fig:geodesics_ellis_wormhole}, and \ref{fig:geodesics_metrictheory_wormhole}, exhibit an excellent agreement with the exact solutions obtained by solving the orbital equation (\ref{eq:orbitalequation}). The main and only limitation of our approach is related to  critical and quasi-critical geodesics due to the instability of the photon trajectories owed to the discrete nature of our algorithm. We obtained an angular deviation lower than $3^\circ$ for both, the Schwarzschild BH and the Ellis WH under the simulation conditions of this work (see Figs. \ref{fig:geodesics_schwarzschild_black_hole} and \ref{fig:geodesics_ellis_wormhole}). For the WH framed in the metric theory of gravity, we obtained considerably larger errors for quasi-critical geodesics. In general, for geodesics with $b>b_c$, the exact and approximated solutions exhibited a reasonable agreement by presenting low angular deviations. This implies that the algorithm may be exploited for the study of emission lines from accretion disks. The accuracy can be improved by either increasing the resolution of the geodesics (but this naturally entails also as a consequence an increment of the computational times) or  by implementing alternative strategies for estimating the displacement $\Delta \mu_i$ at every step. For photons falling toward a BH singularity, the approximation of the intersection between the geodesics and the BH event horizon is not critical; instead for photons traversing a WH mouth, the way in which the intersection point between the photon orbit and the WH mouth in the universe 1 is computed might give rise to strong deviations in the final ends of the geodesic segment in the universe 2. The isotropic refractive index attains higher values around the WH mouth, therefore there more precision in the calculations should be employed in order to avoid these problematic issues. 

The use of isotropic coordinates is extremely important not only for calculation purposes, but they are also able to intuitively capture the main properties of curved space-times. In addition, this algorithm can be straightforwardly generalized to non-zero EM wavelength cases by taking into account the diffraction effects through full-wave numerical solvers based for example
on the Finite Element Method (FEM) \cite{amor2021adaptive} or the Finite-Difference Time-Domain (FDTD) method \cite{taflove2005computational}.

This algorithm can be applied in the numerical image generation of compact objects such as BHs, neutron stars, and WHs. The results presented in this work are very promising if compared with other established and accepted numerical simulations reported in the literature. The advantages and future potentiality of this approach can be shortly summarized in the following points: (1) investigating gravity and gravitational effects through a model easily buildable in laboratory in order to project tests of gravity in small scales, as well as to simulate also and probe gravitational effects occurring around compact objects; (2) vice versa, gravity can be of inspiration for developing new EM technologies, which can be widely employed for engineering applications and to improve the everyday life of the society.

\begin{acknowledgement}
The authors thanks to Comunidad de Madrid MARTINLARA Project (ref. P2018/NMT-4333). V. D. F. acknowledges Istituto Nazionale di Fisica Nucleare (INFN), Sezione di Napoli, \textit{iniziative specifiche} TEONGRAV, and Gruppo Nazionale di Fisica Matematica of Istituto Nazionale di Alta Matematica for the support. 
\end{acknowledgement}

\begin{dataStatement}
 Data underlying the results presented in this paper are not publicly available at this time but may be obtained from the authors upon reasonable request.
\end{dataStatement}

\begin{oaccess}
This article is licensed under a Creative Commons Attribution 4.0 International License, which permits use, sharing, adaptation,
distribution and reproduction in any medium or format, as long as you give appropriate credit to the original author(s) and the source, provide a link to the Creative Commons licence, and indicate if changes were made. The images or other third party material in this article are included in the article’s Creative Commons licence, unless indicated otherwise in a credit line to the material. If material is not included in the article’s Creative Commons licence and your intended use is not permitted by statutory regulation or exceeds the permitted use, you will need to obtain permission directly from the copyright holder. To view a copy of this licence, visit \href{http://creativecommons.org/licenses/by/4.0/}{http://creativecommons.org/licenses/by/4.0/}.\\ Funded by $\text{SCOAP}^\text{3}$. $\text{SCOAP}^\text{3}$ supports the goals of the International Year of Basic Sciences for Sustainable Development.
\end{oaccess}

\bibliographystyle{spphys}       

\bibliography{main.bib}

\begin{thebibliography}{10}
\providecommand{\url}[1]{{#1}}
\providecommand{\urlprefix}{URL }
\expandafter\ifx\csname urlstyle\endcsname\relax
  \providecommand{\doi}[1]{DOI \discretionary{}{}{}#1}\else
  \providecommand{\doi}{DOI \discretionary{}{}{}\begingroup
  \urlstyle{rm}\Url}\fi

\bibitem{eddington1987space}
A.S. Eddington, \emph{Space, time and gravitation: An outline of the general
  relativity theory} (Cambridge University Press, 1987)

\bibitem{plebanski1960electromagnetic}
J.~Plebanski, Physical Review \textbf{118}(5), 1396 (1960)

\bibitem{EHC20191}
K.~Akiyama, et~al., Astrophys. J. Lett. \textbf{875}, L1 (2019).
\newblock \doi{10.3847/2041-8213/ab0ec7}

\bibitem{EHC20194}
K.~Akiyama, et~al., Astrophys. J. Lett. \textbf{875}(1), L4 (2019).
\newblock \doi{10.3847/2041-8213/ab0e85}

\bibitem{EHC20195}
{Event Horizon Telescope Collaboration}, K.~{Akiyama}, et~al., ApJ
  \textbf{875}(1), L5 (2019).
\newblock \doi{10.3847/2041-8213/ab0f43}

\bibitem{EHC20196}
{Event Horizon Telescope Collaboration}, K.~{Akiyama}, et~al., ApJ
  \textbf{875}(1), L6 (2019).
\newblock \doi{10.3847/2041-8213/ab1141}

\bibitem{Akiyama2022image}
K.~Akiyama, et~al., Astrophys. J. Lett. \textbf{930}(2), L14 (2022).
\newblock \doi{10.3847/2041-8213/ac6429}

\bibitem{luminet1979image}
J.P. Luminet, Astronomy and Astrophysics \textbf{75}, 228 (1979)

\bibitem{beloborodov2002gravitational}
A.M. Beloborodov, The Astrophysical Journal \textbf{566}(2), L85 (2002)

\bibitem{placa2019approximation}
R.~La~Placa, P.~Bakala, L.~Stella, M.~Falanga, arXiv preprint arXiv:1907.11786
  (2019)

\bibitem{Poutanen2020}
J.~{Poutanen}, A\&A \textbf{640}, A24 (2020).
\newblock \doi{10.1051/0004-6361/202037471}

\bibitem{vittorio2016approximate}
V.~De~Falco, M.~Falanga, L.~Stella, Astronomy \& Astrophysics \textbf{595}, A38
  (2016)

\bibitem{defalco2021testing}
V.~De~Falco, E.~Battista, S.~Capozziello, M.~De~Laurentis, Physical Review D
  \textbf{103}(4), 044007 (2021)

\bibitem{bao1994emission}
G.~Bao, P.~Hadrava, E.~Ostgaard, The Astrophysical Journal \textbf{435}, 55
  (1994)

\bibitem{aldi2017relativistic}
G.F. Aldi, V.~Bozza, Journal of Cosmology and Astroparticle Physics
  \textbf{2017}(02), 033 (2017)

\bibitem{johannsen2010testingI}
T.~Johannsen, D.~Psaltis, The Astrophysical Journal \textbf{716}(1), 187 (2010)

\bibitem{johannsen2010testingII}
T.~Johannsen, D.~Psaltis, The Astrophysical Journal \textbf{718}(1), 446 (2010)

\bibitem{bakala2007extreme}
P.~Bakala, P.~{\v{C}}erm{\'a}k, S.~Hled{\'\i}k, Z.~Stuchl{\'\i}k,
  K.~Truparov{\'a}, Open Physics \textbf{5}(4), 599 (2007)

\bibitem{fernandez2016anisotropic}
I.~Fern{\'a}ndez-N{\'u}{\~n}ez, O.~Bulashenko, Physics Letters A
  \textbf{380}(1-2), 1 (2016)

\bibitem{falcon2022analogous}
E.~Falc{\'o}n-G{\'o}mez, G.~Santamar{\'\i}a-Botello, V.~De~Falco,
  A.~Amor-Mart{\'\i}n, V.~de~la Rubia, L.E.G. Mu{\~n}oz, in \emph{2021 51st
  European Microwave Conference (EuMC)} (IEEE, 2022), pp. 630--633

\bibitem{weinberg1972gravitation}
S.~Weinberg, \emph{Gravitation and cosmology: principles and applications of
  the general theory of relativity} (1972)

\bibitem{defalco2019coupling}
V.D. Falco, Coupling poynting-robertson effect in mass accretion flow physics.
\newblock Ph.D. thesis, Universitat Basel (2019)

\bibitem{kunz1954propagation}
K.~Kunz, Journal of Applied Physics \textbf{25}(5), 642 (1954)

\bibitem{carroll2019spacetime}
S.M. Carroll, \emph{Spacetime and geometry} (Cambridge University Press, 2019)

\bibitem{morris1988wormholes}
M.S. Morris, K.S. Thorne, American Journal of Physics \textbf{56}(5), 395
  (1988)

\bibitem{bogush2022photon}
I.~Bogush, D.~Gal'tsov, G.~Gyulchev, K.~Kobialko, P.~Nedkova, T.~Vetsov, Phys.
  Rev. D \textbf{106}, 024034 (2022).
\newblock \doi{10.1103/PhysRevD.106.024034}.
\newblock \urlprefix\url{https://link.aps.org/doi/10.1103/PhysRevD.106.024034}

\bibitem{page1974disk}
D.N. Page, K.S. Thorne, The Astrophysical Journal \textbf{191}, 499 (1974)

\bibitem{perlick2022calculating}
V.~Perlick, O.Y. Tsupko, Physics Reports \textbf{947}, 1 (2022)

\bibitem{DNGimages}
Textures used in the interstellar movie by dneg. online:
  \href{https://www.dneg.com/interstellar-wormhole/}{https://www.dneg.com/interstellar-wormhole/}
  (2015).
\newblock \urlprefix\url{https://www.dneg.com/interstellar-wormhole/}

\bibitem{amor2021adaptive}
A.~Amor-Martin, L.E. Garcia-Castillo, Applied Sciences \textbf{11}(8), 3683
  (2021)

\bibitem{taflove2005computational}
A.~Taflove, S.C. Hagness, M.~Piket-May, The Electrical Engineering Handbook
  \textbf{3}, 629 (2005)

\end{thebibliography}

\end{document}